# The Solar Cause of the 2022 February 3 Geomagnetic Storm that Led to the Demise of the Starlink Satellites

Nat Gopalswamy[1], Hong Xie[1,2], Seiji Yashiro[1,2], and Sachiko Akiyama[1,2]

[1]NASA Goddard Space Flight Center, Greenbelt, MD 20771, USA

[2]The Catholic University of America, Washington DC 20064, USA

e-mail: nat.gopalswamy@nasa.gov

**Abstract:**

We report on the solar source of the 2022 February 3 geomagnetic storm of moderate strength that contributed to the loss of 39 Starlink satellites. The geomagnetic storm was caused by the 2022 January 29 halo coronal mass ejection (CME) that was of moderate speed (~690 km/s) originating from NOAA active region 12936 located in the northeast quadrant (N18E06) of the Sun. The eruption was marked by an M1.1 flare, which started at 22:45 UT, peaked at 23:32 UT on January 29 and ended at 00:24 UT the next day. The CME ended up as a shock-driving magnetic cloud (MC) observed at Sun-Earth L1 and at STEREO-Ahead (STA) located ~34⁰ behind Earth. The geomagnetic storm was caused by a strong southward component of the MC that was boosted by a high speed solar wind stream behind the MC. Even though Earth and STA were separated by only ~34⁰, the MC appeared quite different at Earth and L1. One possibility is that the MC was writhed reflecting the curved neutral line at the Sun. In-situ observations suggest that the MC was heading closer to STA than to Earth because of the earlier arrival at STA. However, the shock arrived at STA and Earth around the same time, suggesting a weaker shock at Earth due to flank passage.

## 1. Introduction

Space Weather affects humans and their technology in space and on ground. Space weather in Earth's thermosphere is of concern because of satellites in Low Earth Orbit (LEO). A recent example is the Starlink incident in February 2022, during which 38 of the 49 satellites launched by SpaceX's Falcon 9 perished. The launch occurred when a geomagnetic storm was in its recovery phase on February 3, 2022. The geomagnetic storm was of moderate intensity (minimum Dst ~ -66 nT). The geomagnetic storm resulted in an elevated thermospheric density and hence the atmospheric drag over an extended period of time that led to the premature deorbit of the 38 satellites (Dang et al. 2022; Zhang et al. 2022; Fang et al. 2022; Lin et al. 2022). The storm was associated with the 2022 January 29 at 23:36 UT halo coronal mass ejection (CME), which arrived at Earth three days later (Dang et al. 2022).

Several aspects of the Starlink event have already been reported in the above papers, so we focus on the solar source of the underlying halo CME, the early CME kinematics, and how the CME ended up as a magnetic cloud (MC) at 1 AU. We identified the MC using the criteria established by Burlaga et al. (1981): enhanced magnetic field strength, smooth rotation of one of the magnetic field components, and low plasma beta.

The storm was caused by a CME, which is a bundle of magnetic field lines embedded in a hot plasma moving away from the Sun with speeds up to 3000 km/s. The CME left the Sun on January 29 at 23:58 UT and arrived at Earth about 3 days later. The magnetic field of the CME interacted with Earth's magnetic field resulting in the geomagnetic storm. The CME was of moderate speed (~550 km/s), but fast enough to drive a shock that caused a small increase in Dst (known as storm sudden commencement) preceding the geomagnetic storm.

## 2. Observations, analysis, and results

The eruption occurred in NOAA active region (AR) 12936, which was located in the northeast quadrant of the Sun (N17E11). The active region was mainly bipolar with leading positive polarity as observed by the Helioseismic Magnetic Imager (HMI, Scherrer et al. 2012) on board the Solar Dynamics Observatory (SDO, Pesnell et al. 2012). Various eruption signatures were observed in EUV images obtained by SDO's Atmospheric Imaging Assembly (AIA, Lemen et al. 2012) at several wavelengths. Additional EUV images were also obtained by the Extreme Ultraviolet Imager (EUVI), which is part of the Sun Earth Connection Coronal and Heliospheric Investigation (SECCHI, Howard et al. 2008) instrument suit on board the Solar Terrestrial Relations Observatory (STEREO, Kaiser et al. 2008). SECCHI's inner (COR1) and outer (COR2) coronagraphs and the Large Angle and Spectrometric Coronagraph (LASCO, Brueckner et al. 1995) on board the Solar and Heliospheric Observatory (SOHO, Domingo et al. 1995) observed the CME in the corona and inner heliosphere. LASCO's C2 and C3 telescopes cover the corona in the heliocentric distance range 2.5 to 32 solar radii (Rs), while COR1 and COR2 observe over the heliocentric distance range of 1.4 to 15 Rs. At the time of this event, STEREO-Ahead (STA), located at E34 was observing the Sun. Thus, the white light CME in this event is well observed from the Sun-Earth line (SOHO, SDO) and from off the Sun-Earth line. The flare aspects of the eruption was obtained from GOES soft X-ray light curve in addition to the information provided by the EUV imagers. Finally, in-situ observations of the solar wind including the CME are obtained from the OMNI (Operating Missions as a Node) and STEREO's Plasma and Suprathermal Ion Composition (PLASTIC, Galvin et al. 2008) and In situ Measurements of Particles And CME Transients (IMPACT, Luhmann et al. 2008) investigations. The in-situ observations are combined with the coronagraph and EUV observations to infer the Sun-to-Earth propagation of the CME.

### 2.1 The eruption geometry

Figure 1 shows the active region of interest in an HMI magnetogram. A post eruption arcade (PEA) formed straddling the main neutral line in the active region as observed by AIA at 193 Å.

The AIA image also shows dimming regions D1 and D2 located on the negative and positive sides of the polarity inversion line, respectively. The dimming regions are thought to be the locations where the legs of the CME flux rope are anchored (e.g., Webb et al. 2000; Gopalswamy 2009; Dissauer et al. 2018; Sindhuja and Gopalswamy 2020). The location of the dimming regions suggests that the axis of the CME flux rope is roughly horizontal and the axial magnetic field points to the east. The azimuthal field is north pointing in front of the axis and south pointing in the back. The inferred direction of the flux rope axial magnetic field suggests that the flux rope has a right handed helicity, which typically occurs in the southern hemisphere. Our active region is in the northern hemisphere, so the flux rope violates the hemispheric helicity sign rule. Such violations are not uncommon (Zhang et al. 2016). The PEA is consistent with the flare ribbons observed in SDO/AIA 1600 Å images shown in Figure 2.

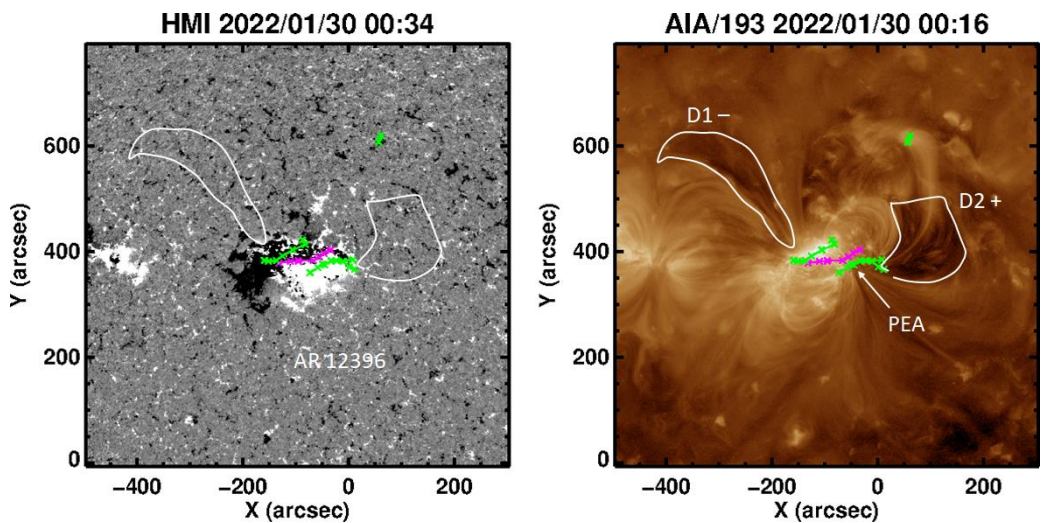

Figure 1. The eruption geometry revealed by the SDO/HMI magnetogram (left) and an EUV image from SDO/AIA at 193 Å (right). The dimming regions on either side of the polarity inversion line (pink line) are denoted by D1 and D2 that have negative and positive polarities at the photospheric level. The green crosses mark the feet of the post eruption arcade (PEA).

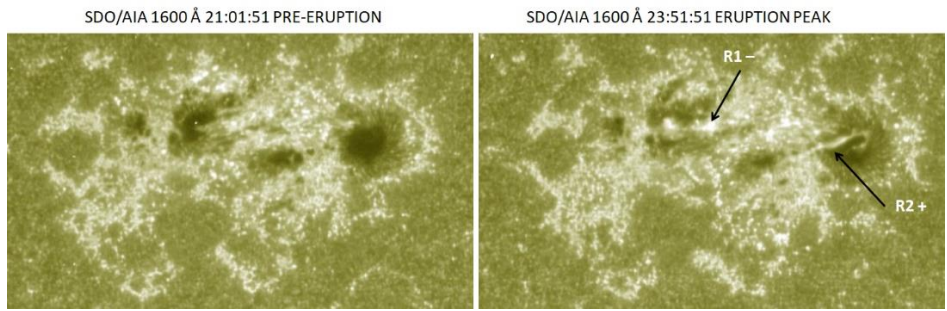

Figure 2. AR 12936 with sunspots as seen in SDO/AIA 1600 Å mages taken at 21:01:51 and 23:51:51 UT. The presence of the flare ribbons is evident in the right image. R1 and R2 are ribbons on the negative and positive polarity patches of the active region. Both the ribbons extend into the sunspots.

GOES soft X-ray light curve in Figure 3 shows that the flare is of M-class (M1.1), which started, peaked, and ended at 22:45 UT, 23:32 UT and 00:24 UT, respectively (the end time is on January 30). The flare was accompanied by a CME, which was well observed by LASCO and SECCHI coronagraphs. The CME first appeared in the field of view (FOV) of STA/COR1 at 23:16:35 UT (https://cor1.gsfc.nasa.gov/catalog/cme/2022/html/20220129_2311_cor1.html) and in the LASCO/C2 FOV at 23:36 UT (https://cdaw.gsfc.nasa.gov/movie/make_javamovie.php?img1=lasc2rdf&stime=20220129_2211&etime=20220130_0311). When the CME first appeared above the northeast limb, it was not a halo, but in the next frame it became a halo CME (Howard et al. 1982; Gopalswamy et al. 2010a). Figure 3 shows the SOHO/LASCO CME at two instances. Detailed information on the CME and the associated phenomena can be found in the SOHO/LASCO CME catalog https://cdaw.gsfc.nasa.gov/CME_list/UNIVERSAL/2022_01/univ2022_01.html (Yashiro et al. 2004; Gopalswamy et al. 2009). In the sky plane, the CME leading edge has a speed of ~530 km/s slowly decelerating at the rate of -10.1 m s$^{-2}$.

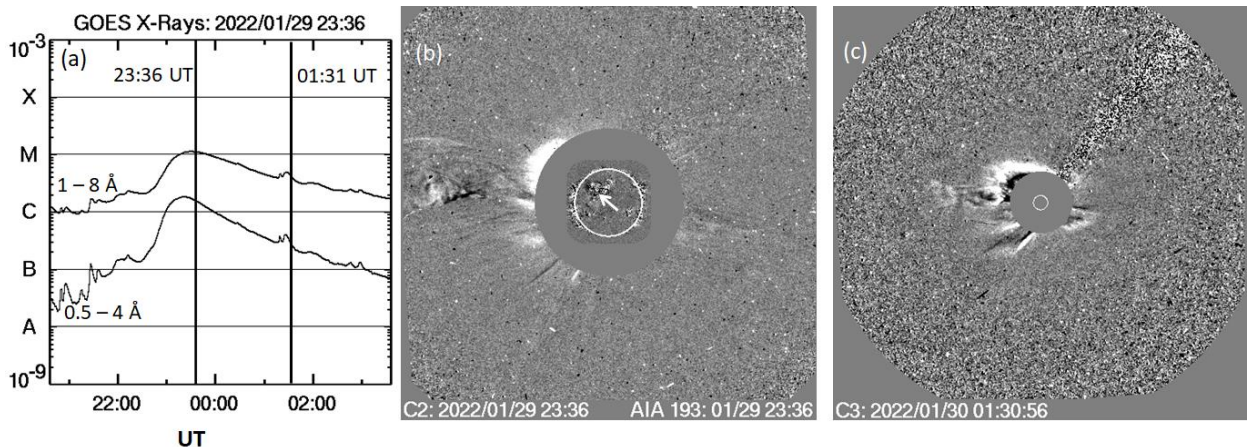

Figure 3. GOES soft X-ray light curve (W m$^{-2}$) showing the M1.1 flare (a), along with SOHO/LASCO difference images at 23:36 UT (b) and 01:31 UT (c). The times of the images are marked on the GOES light curve in (a) by the vertical solid lines. In (b) and (c) the small white circles at the center represent the optical Sun. In (b), an EUV difference image at 193 Å taken around the same time as the LASCO/C2 image is superposed. The disturbances pointed to by an arrow in (b) identifies the eruption region on the solar disk.

## 2.2 CME kinematics

The sky-plane speed is typically smaller than the actual speed of the CME. The actual speed can be determined using a flux rope fit to the LASCO and STEREO images of the CME. Here we use the Elliptical Flux Rope (EFR) model, which assumes that the CME flux rope has an elliptical axis with varying radial circular cross-section. The tip of the flux rope is at a heliocentric distance $h_{FR}$. The radius ($R_0$) of the flux rope at its apex is related to the axial distance ($h_{FR} – R_0$) via the geometrical parameter $\Lambda = (h_{FR} – R_0)/2R_0$. From $\Lambda$ we obtain the flux rope radius as $R_0 = h_{FR}/(2\Lambda + 1)$. Full details of the model can be found in Krall and St Cyr (2006) and Krall (2007). Figure 4 shows the STA/COR2 image at 00:30 UT on January 30 with the fitted EFR flux rope superposed. The fit is reasonable, yielding $R_0/h_{FR} = 0.36$, which corresponds to $\Lambda = 0.89$. This solution indicates that when the flux rope leading edge is at 10 Rs, the flux rope has a radius of 3.6 Rs. The flux rope axis is mostly horizontal and the tilt angle is rather small (-18º). This is consistent with the locations of the dimming regions D1 and D2 shown in Figure 1. The EFR fit also determines the edge-on and face-on widths of the flux rope as 29º and 45º, respectively.

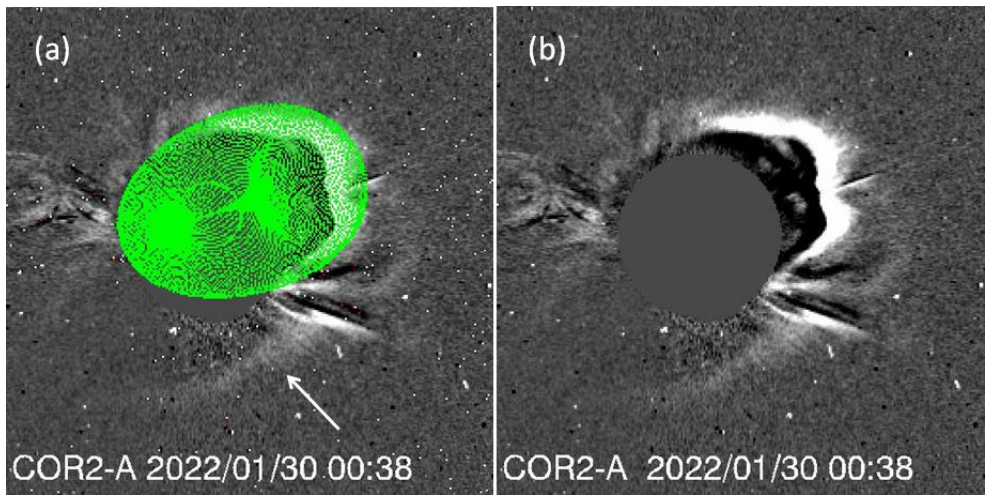

Figure 4. (a) The flux rope from the EFR fit (green mesh) overlaid on the STA/COR2 image at 00:38 UT. The white arrow points to the disturbance (likely to be a weak shock) surrounding the flux rope (b) The STA/COR2 image at 00:38 UT shown without the EFR fit for reference.

Tracking the leading edge of the flux rope, we determined the height-time history of the flux rope, its speed, and acceleration profiles as shown in Figure 5. The CME speed slowly increases and reaches a peak speed of ~ 744 km/s just before the initial acceleration ends. The slow increase in speed is consistent with the low peak acceleration of ~ 357 m s$^{-2}$. The initial acceleration can also be obtained from the flare rise time (~47 min) and the maximum CME speed (744 km/s) attained as 260 m s$^{-2}$. This is an acceleration averaged over the flare rise time and hence smaller than the peak acceleration. After attaining the peak speed, the CME slowly decelerated at the rate of -7.1 m s$^{-2}$. The average speed of the flux rope obtained from a linear fit to the EFR data points in the LASCO FOV is ~691 km/s, significantly larger than the sky plane speed (530 km/s). The CME continued in the FOV of the STEREO's Heliospheric Imager (HI, Howard et al. 2008) and eventually observed in situ by spacecraft at L1 and STA. The CME appeared in the HI-1 FOV at 2:48 UT on January 30. More details on the CME observed in the HI FOV can be found in:

https://www.helcats-fp7.eu/catalogues/event_page.html?id=HCME_A__20220130_01.

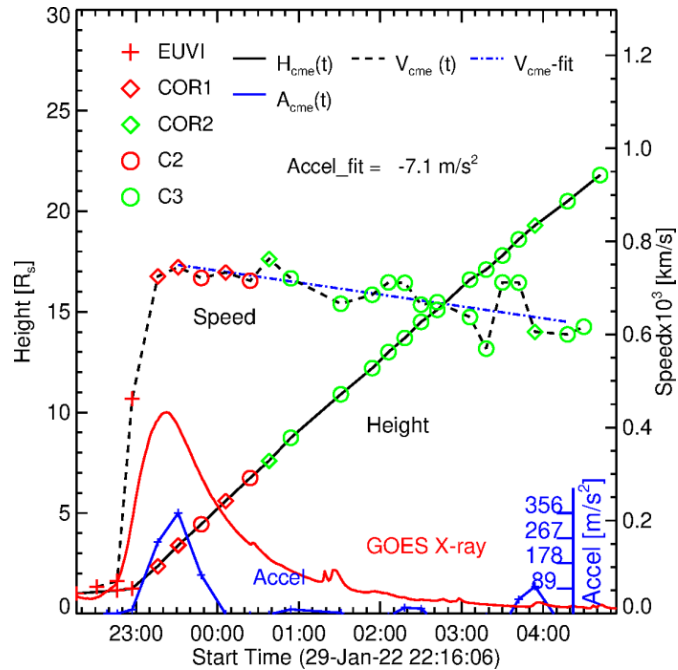

Figure 5. EFR leading edge height ($H_{cme}(t)$), speed ($V_{cme}(t)$), and acceleration ($A_{cme}(t)$) plotted as a function of time using images from STA's EUVI, COR1, and COR2, and SOHO/LASCO's C2 and C3. These data points are distinguished on the plot using different symbols and colors. The CME peak speed (744 km/s) is attained around the time of the flare maximum and slowly decayed after that at the rate of -7.1 m s$^{-2}$.

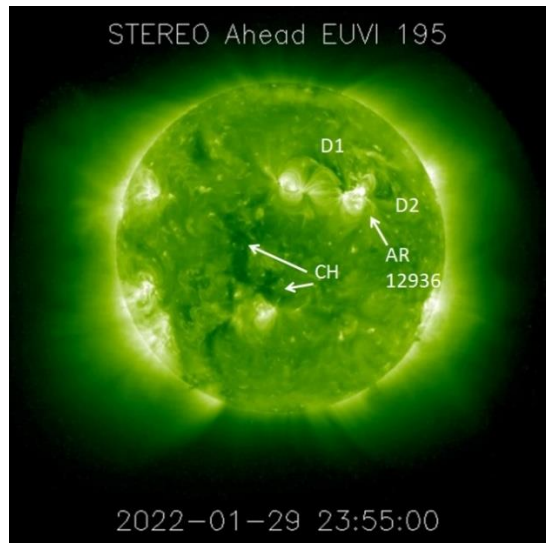

Figure 6. Coronal environment of the eruption region (AR 12936) as viewed from an STA/EUVI 195 Å image when the eruption was in progress. The dimming regions D1 and D2 and a coronal hole (CH) of interest are marked on the plot. The bright structure in the active region is the PEA (see Figure 1).

Dang et al. (2022) fit a graduated cylindrical shell (GCS, Thernisien 2011) fit to the CME flux rope. Our EFR fit generally agrees with the GCS fit, except for the flux rope propagation direction. While the flare location is N17E11, the EFR fit gives a flux rope location of N05E20, slightly to the east and south of the flare location. On the other hand the GCS fit gives a flux rope location of S07E34. The typical errors of the GCS fit are ±17⁰ in longitude and ±4⁰ in latitude (Thernisien et al. (2009) with similar numbers for the EFR fit. Therefore, the southward shift is significant for both the fits, while eastward shift is significant for the GCS fit. There do not seem to be any possible sources of deflection to the north and west of the eruption region, except for the north polar coronal hole. Deflection due to the magnetic field in the polar coronal hole in the rise phase of the solar cycle is well known (Gopalswamy et al. 2008). There are coronal holes to the southeast of the source active region (see Figure 6) that would have partially offset the southward deflection. The southeastern coronal hole is likely to have resulted in a high speed stream causing a tailwind to the CME flux rope. The different location of the GCS fit is probably due to the inclusion of the surrounding disturbances as part of the flux rope (see Figure 4).

We measured the total reconnected (RC) flux in the eruption region using the line of sight magnetogram and the EUV images of the PEA (Gopalswamy et al. 2017): the total RC flux is half the magnetic flux in the area underlying the PEA. We chose the AIA 193 Å image at 01:30 UT to determine the area underlying the PEA. The total RC flux was estimated as $2.55 \times 10^{21}$ Mx. Using a set of more than 50 ICMEs, Gopalswamy et al. (2018a) obtained an empirical relation between the CME flux rope speed ($V$) and total RC flux ($\varphi$) as:
$V = 394 \, \varphi^{0.67}$,   (1)
where $\varphi$ is in units of $10^{21}$ Mx. For $\varphi = 2.55$, Eq. (1) gives $V = 738$ km/s, in close agreement with the measured flux rope speed (691 km/s).

Using the Flux Rope from Eruption Data (FRED) technique (Gopalswamy et al. 2018b; Sarkar et al. 2020), it is possible to estimate the axial magnetic field of the near-Sun flux rope, say at a distance of 10 Rs. The technique is to equate $\varphi$ to the poloidal flux ($\varphi_p$) of the CME flux rope derived from forward fitting, so that the axial field strength ($B_o$) can be determined:
$B_o = \varphi_p \, x_{01} / R_0 L$,    (2)
where $x_{01}$ (=2.4048) is the first zero of the Bessel function $J_0$, $R_0$ is the flux rope cross-sectional radius, and $L$ is the flux rope length obtained from the flux-rope height ($h_{FR} - R_0$) and the angular width from the EFR fit ($2 \times 74^\circ$ or 2.58 radians). When $h_{FR} = 10$ Rs, the axis is at a distance of 6.4 Rs, so $L = 16.5$ Rs. Substituting these values in (2), and using $\varphi_p = 2.55 \times 10^{21}$ Mx, we get an axial field strength of 21.3 mG. This value is at the lower end of the axial field strength distribution of coronal flux ropes (average ~51.9 mG) and is appropriate for the low measured RC flux.

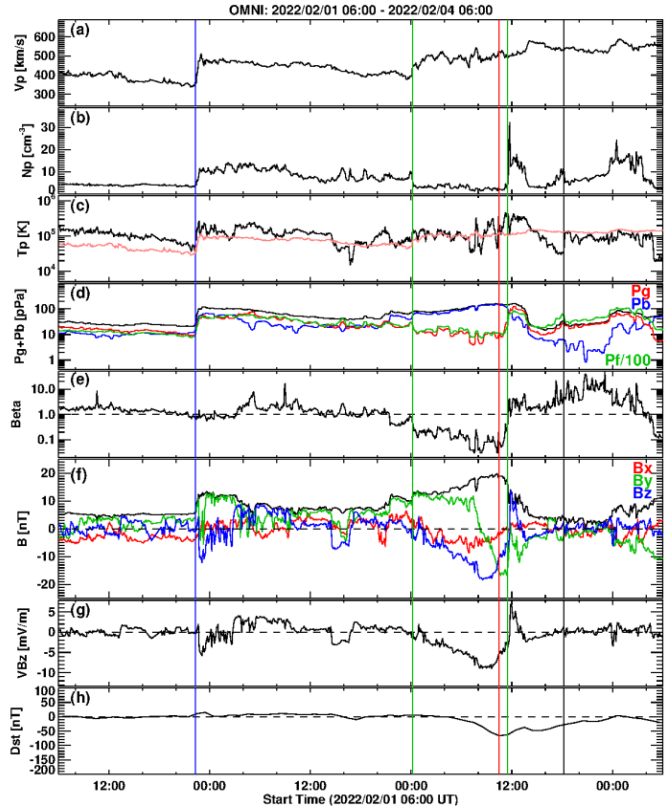

Figure 7. Solar wind observations from OMNI for the period 2022 February 1 – 4. (a) Solar wind speed ($V_p$), (b) proton density ($N_p$), (c) proton temperature ($T_p$) along with the expected temperature (pink line), (d) gas ($P_g$ – red curve), magnetic ($P_b$ – blue curve), and flow ($P_f$ – green curve) pressures and the total pressure ($P_g+P_b$ – black curve), (e) plasma beta, (f) total magnetic field strength (B) along with the three components $B_x$ (red curve), $B_y$ (green curve), and $B_z$ (blue curve) in GSE coordinates, (g) solar wind electric field (solar wind speed times the $B_z$ component of the magnetic field), (h) the $D_{st}$ index showing the moderate geomagnetic storm (time of Dst minimum marked by the vertical red line: 11:00 UT on February 3). The $D_{st}$ data are from the World Data Center, Kyoto. The vertical green lines mark the boundaries of the magnetic cloud based on plasma low beta signature. The vertical blue line marks the arrival of the weak shock driven by the MC. The vertical black line marks the launch time of the Starlink satellites (18:13 UT on February 3).

## 2.3 Interplanetary CME

Figure 7 shows the solar wind parameters from the OMNI data base. The ICME is accompanied by a shock that arrives at the Wind spacecraft at 21:27 UT on February 1 (https://lweb.cfa.harvard.edu/shocks/wi_data/00791/wi_00791.html). The shock was also detected by SOHO's Proton Monitor (Ipavich et al. 1998) at 21:42 UT as can be found in the SOHO shock list: https://space.umd.edu/pm/figs/figs.html. Earth arrival was ~54 minute later. Thus, the shock arrival at Earth corresponds to a transit time of ~67 hours counting from the first appearance time of the CME in LASCO C2 FOV. The shock sheath lasts for about a day because the shock is relatively weak. The MC-type ICME arrives at ~0:10 UT on February 3 and lasts for ~11.3 hrs. The speed profile does not show the typical flux rope expansion (Figure 7a), but increases toward the back of the MC. This is indicative of compression of the MC, as evidenced by the density increase immediately following the MC (Figure 7b). The proton temperature also does not show the typical low temperature signature (Figure 7c). The total pressure inside the MC (Figure 7d) is dominated by the magnetic pressure ($P_b$). The pressure is higher at the end of the cloud indicating compression of the MC. The interval of high $P_b$ is coincident with the low plasma beta (Figure 7e) as expected. The MC compression is evident from the sudden increase in the Bz component by ~50% (from -12 nT to -18 nT, see Figure 7f) towards the end of the cloud. The dynamic pressure inside the MC ($P_f$) is relatively low, so the Dst index is mainly controlled by the reconnection electric field, VBz (see Figure 7g). The Dst index attains its minimum value immediately after the Bz depression (Figure 7h). The Bz component sharply turns northward after the MC, resulting in a rapid, brief recovery of Dst followed by a normal recovery. The MC has its $B_z$ mostly negative indicating it is a high-inclination cloud as observed at end. The By component mostly in the eastward direction, so this is a right-handed MC.

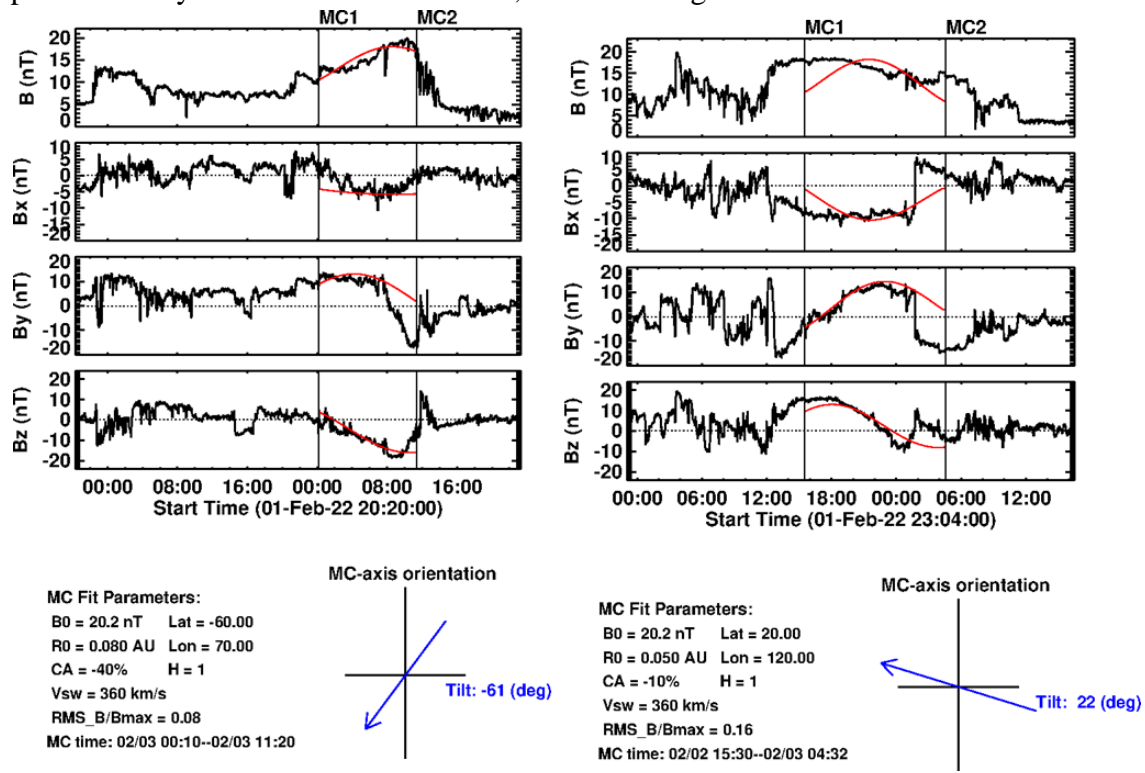

Figure 8. Comparison between in situ magnetic field observations at Wind (OMNI, left) and STA (right) along with the fitted parameters (red lines). The blue arrows at the bottom show the local orientation of the flux rope axis at Wind and STA. The fit parameters shown at the bottom are: the axial field strength Bo, MC radius Ro, closest approach parameter (CA) in percentage, helicity sign (H), solar wind speed (Vsw), the RMS magnetic field fluctuation relative to the maximum field strength (B/Bmax).

The launch of the Starlink satellites occurred during the recovery phase of the storm, but the storm-induced density in the thermosphere was significantly elevated during the recovery phase as measured by the SWARM satellites (Lin et al. 2022). The density increase was further boosted by another geomagnetic storm that had a minimum Dst of -61 nT on February 4 at 21 UT. Details on the density variability during the two storms can be found in Zhang et al. (2022).

From the EFR fit, we infer that the flux rope is heading along E20, which is slightly closer to STA (E34) than to Earth (E00). This is consistent with the earlier arrival of the MC at STA by ~8.5 hrs (see Figure 8 for a comparison of the magnetic field components observed at Wind and STA). The fits also indicate that the closest approach parameter (CA) is smaller for STA than for Wind indicating STA passage is closer to the MC axis. The shock arrivals, on the other hand, are much closer at the two spacecraft: 00:30 UT (February 2) at STA compared to 21:30 UT (February 1) at Wind. WSA-Enlil simulations performed by Fang et al. (2022) do confirm this shock arrival time at Earth. The standoff distance is also much longer at Wind indicated by the longer-duration sheath (23 hrs vs. 15 hrs at STA). These observations suggest that the flux rope nose is closer to STA than at Wind and the weaker shock at Wind stands at a larger distance. The flux rope appears different at Wind and STA with different tilts. We speculate that the flux might be distorted (possible writhing) by the time it is observed in situ. This needs further investigation. We also notice that the flux rope radius was not expanding self similarly because of the interaction with the high speed stream.

## 3. Discussion

The 2022 February 3 MC was driving a shock at 1 AU, but there was no type II radio burst reported by STEREO/WAVES (Bougeret et al. 2008) or by ground based observatories. Such radio quiet shocks are well known and are not uncommon (Gopalswamy et al. 2010b). We did observe shock-like disturbances surrounding the flux rope, but the shock was not strong enough to accelerate nonthermal electrons to produce a type II radio burst. There was a metric nonthermal radio continuum starting around 23:06 UT and ending early next day (00:15 UT) in the frequency range 25-80 MHz. The radio emission was recorded by the Compound Astronomical Low frequency Low cost Instrument for Spectroscopy and Transportable Observatory (CALLISTO, Benz et al. 2005) spectrometer located at Cohoe, Alaska (http://soleil.i4ds.ch/solarradio/callistoQuicklooks/?date=20220129). The eruption was also associated with a low-frequency type III radio burst observed by the STEREO/WAVES experiment, starting at 23:20 UT below 2 MHz and extending down to 50 kHz (http://stereo.space.umn.edu/data/level-3/STEREO/Both/SWAVES/daily-summary-plots/gray-status/PNG/2022/stereo_swaves_daily-summary_gray-status_20220130_v04.png). Note that both the flare continuum and the type III burst are produced by electrons accelerated during flare reconnection that access post flare loops and open field lines, respectively.

The shock transit time from the first appearance time in LASCO/C2 (23:36 UT on January 29) to the shock arrival time at STA (00:30 UT on February 2) is ~72 hrs. The empirical Shock Arrival (ESA) model gives the shock transit time T in terms of the average CME speed in the coronagraph FOV (V) according to the relation (Gopalswamy et al. 2005a,b):
$T = 151.002 \times 0.998625^V + 11.5981$ hrs (3).

Recall that the average speed of the CME in the coronagraph FOV is 691 km/s, which when substituted in equation (1) gives T = 69.95 hours, only ~2 hrs shorter than the observed transit time (72 hrs). If we take into account of the fact that the CME longitude is ~14º to the west of the Sun-STA line, the transit time becomes 71.82 hrs, almost the same as the observed value. This suggests that the CME underwent normal deceleration, not significantly affected by the following high speed stream.

## 4. Summary and conclusions

We investigated the solar source of the 2022 February 3 geomagnetic storm that initiated the Starlink space weather event. The CME was of moderate speed with an average speed of ~691 km/s in the coronagraph FOV. The CME was also observed by STEREO/HI in the interplanetary medium before being detected at Sun-Earth L1 by Wind and SOHO spacecrafts. Three-dimensional information of the CME flux rope obtained from the EFR fitting indicates that the CME was heading in a direction between Earth and STEREO-Ahead but closer to the latter. The main findings of the present study can be summarized as follows.

1. The geomagnetic storm in question is caused by the southward magnetic field component of the magnetic cloud that arrived at Earth at the beginning of February 3, 2022.

2. Due to high speed wind that followed the magnetic cloud, the southward magnetic field component was enhanced significantly just before the Dst attained its minimum value.

3. The magnetic cloud can be traced back to the Sun as a halo CME originating from NOAA active region 12936 that erupted on January 29, 2022 at 22:45 UT.

4. The halo CME is of moderate speed (~690 km/s) in the coronagraph FOV and slowly decelerated in the interplanetary medium. The initial acceleration peaked at 0.36 km s$^{-2}$, which is consistent with the average acceleration derived from flare rise time and CME speed.

5. The CME speed is consistent with the empirical relation between CME speed and total reconnected flux derived from the magnetic flux underlying the post eruption arcade.

6. The flux rope radius and the axial magnetic field strength at a distance of ~10 Rs are within the appropriate ranges obtained from statistical results.

7. While the CME-driven shock arrived at STEREO-Ahead and Earth within two hours, the magnetic cloud arrived at STEREO-Ahead some 15 hours ahead, suggesting flank arrival at Earth.

8. The MC flank arrival is confirmed by the large sheath thickness at Earth as compared to that at STEREO-Ahead.

**Acknowledgments**

This work benefited from NASA's open data policy. We thank the SOHO, STEREO, SDO, and Wind teams for making their data available on-line that are utilized in the present investigation. We also acknowledge the use of GOES X-ray data made available online by NOAA. Work supported by NASA's LWS TR&T and STEREO programs. We thank the referee A. Klicik for helpful comments.

**References**

Benz, A. O., Monstein, C., Meyer, H.: 2005, Solar Phys. 226, 143. DOI: 10.1007/s11207-005-5688-9

Bougeret, J.-L., Goetz, K., Kaiser, M. L. et al.: 2008, Space Sci. Rev. 136, 487. DOI: 10.1007/s11214-007-9298-8

Brueckner, G. E., Howard, R. A., Koomen, M. J. et al.: 1995, Solar Phys. 162, 357. DOI: 10.1007/BF00733434

Burlaga, L.F., Sittler, E., Mariani, F., Schwenn, R.: 1981, J. Geophys. Res. 86, 6673. DOI: 10.1029/JA086iA08p06673

Dang, T., Li, X., Luo, B., Li, R.,Zhang, B., Pham, K., et al.: 2022, Space Weather, 20, e2022SW003152. DOI: 10.1029/2022SW003152

Dissauer, K., Veronig, A. M., Temmer, M., Podladchikova, T., and Vanninathan, K.: 2018, Astrophys. J., 855, article id.137. DOI: 10.3847/1538-4357/aaadb5

Domingo, V., Fleck, B., Poland, A. I.: 1995, Solar Phys. 162, 1. DOI: 10.1007/BF00733425

Fang, T.-W., Kubaryk, A., Goldstein, D., Li, Z., Fuller-Rowell, T., Millward, G., et al.: 2022, Space Weather, 20, e2022SW003193. DOI: 10.1029/2022SW003193

Galvin, A. B., L. M. Kistler, M. A. Popecki et al.: 2008, Space Sci. Rev. 136, 437. DOI: 10.1007/s11214-007-9296-x

Gopalswamy, N.: 2009, Climate and Weather of the Sun-Earth System (CAWSES): Selected Paper from the 2007 Kyoto Symposium, Ed. by T. Tsuda, R. Fujiki, K. Shibata, and M. A. Geller, 77.

Gopalswamy, N., Lara, A., Manoharan, P. K., Howard, R. A.: 2005a, Adv. Space Res. 36, 2289. DOI: 10.1016/j.asr.2004.07.014

Gopalswamy, N., Yashiro, S., Liu, Y., Michalek, G., Vourlidas, A., Kaiser, M.L., and Howard, R.A.: 2005b, J. Geophys. Res. 110, A09S15. DOI: 10.1029/2004JA010958

Gopalswamy, N., Akiyama, S., Yashiro, S., Michalek, G., Lepping, R. P.: 2008, J. Atmos. Sol. Terr. Phys. 70, 245. DOI: 10.1016/j.jastp.2007.08.070

Gopalswamy, N., Yashiro, S., Michalek, G., Stenborg, G., Vourlidas, A., Freeland, S., and Howard, R. A.: 2009, Earth Moon Planets 104, 295. DOI: 10.1007/s11038-008-9282-7

Gopalswamy, N., Yashiro, S., Michalek, G., Xie, H., Mäkelä, P., Vourlidas, A., and Howard, R. A.: 2010a, Sun and Geosphere. 5, 7.

Gopalswamy, N., Xie, H., Mäkelä, p., Akiyama, S., Yashiro, S., Kaiser, M. L., Howard, R.A. and Bougeret J.-L.: 2010b, Astrophys. J. 710, 1111. DOI: 10.1088/0004-6256/710/2/1111

Gopalswamy, N., Yashiro, S., Akiyama, S., and Xie, H.: 2017, Solar Phys. 292, 65. DOI: 10.1007/s11207-017-1080-9

Gopalswamy, N., Yashiro, S., Akiyama, S., and Xie, H.: 2018a, J. Atmos. Sol. Terr. Phys. 180, 35. DOI: 10.1016/j.jastp.2017.06.004

Gopalswamy, N., Yashiro, S., Akiyama, S., and Xie, H.: 2018b, IAU Symposia, 335, 258. DOI: 10.1017/S1743921317011048

Howard, R. A., Michels, D. J., Sheeley, N. R. Jr. and Koomen, M. J.: 1982, Astrophys. J. 263, L101. DOI: 10.1086/183932

Howard, R. A., Moses, J. D., Vourlidas, A. et al.: 2008, Space Sci. Rev. 136, 67. DOI: 10.1007/s11214-008-9341-4

Ipavich, F. M., Galvin, A. B., Lasley, S. E. et al.: 1998, J. Geophys. Res. 103, 17205. DOI: 10.1029/97JA02770

Kaiser, M. L., Kucera, T. A., Davila, J. M., St. Cyr, O.C., Guhathakurta, M., Christian, E.: 2008, Space Sci. Rev. 136, 5. DOI: 10.1007/s11214-007-9277-0

Krall, J.: 2007, Astrophys. J. 657, 559. DOI: 10.1086/510191

Krall, J., and Sr. Cyr, O. C.: 2006, Astrophys. J., 652, 1740. DOI: 10.1086/508337

Lemen, J., Title, A., Akin, D. J. et al.: 2012, Solar Phys. 275, 17. DOI: 10.1007/s11207-011-9776-8

Lin, D., Wang, W., Garcia-Sage, K., Yue, J., Merkin, V., McInerney, J. M., et al.: 2022, Space Weather, 20, e2022SW003254. DOI: 10.1029/2022SW003254

Luhmann, J. G., Curtis, D. W. Schroeder, P. et al.: 2008, Space Sci. Rev. 136, 5. DOI: 10.1007/s11214-007-9170-x

Pesnell, W. D., Thompson, B. J., Chamberlin, P. C., et al.: 2012, Solar Phys. 275, 3. DOI: 10.1007/s11207-011-9841-3

Sarkar, R., Gopalswamy, N., and Srivastava, N.: 2020, Astrophys. J. 888, 121. DOI: 10.1086/508337

Scherrer, P. H., Schou, J., Bush, R. I. et al.: 2012, Solar Phys. 275, 207. DOI: 10.1007/s11207-011-9834-2

Sindhuja, G., and Gopalswamy, N.: 2020, Astrophys. J., 889, id.104. DOI: 10.3847/1538-4357/ab620f

Thernisien, A.: 2011, Astrophys. J. Supp. Ser. 194, 33. DOI:10.1088/0067-0049/194/2/33

Thernisien, A., Vourlidas, A., Howard, R.A.: 2009. Forward modeling of coronal mass ejections using STEREO/SECCHI data. Sol. Phys. 256, 111.

Webb, D. F., Lepping, R. P., Burlaga, L. F., DeForest, C. E., Larson, D. E., Martin, S. F., Plunkett, S. P., and Rust, D. M.: 2000, J. Geophys. Res. 105, 27251. DOI: 10.1029/2000JA000021

Yashiro, S., Gopalswamy, N., Michalek, G., St. Cyr, O. C., Plunkett, S. P., Rich, N. B., and Howard, R. A.: 2004, J. Geophys. Res. 109, A07105. DOI: 10.1029/2003JA010282

Zhang, H., Brandenburg, A., Sokoloff, D. D.: 2016, Astrophys. J. 819, 146. DOI: 10.3847/0004-637X/819/2/146

Zhang, Y., Paxton, L. J., Schaefer, R., & Swartz, W. H.: 2022, Space Weather. 20, e2022SW003168. DOI: 10.1029/2022SW003168